# Constructor theory of time

David Deutsch & Chiara Marletto

Constructor theory is a research programme in fundamental physics, proposing the principle that the laws of physics are expressible as specifications of which transformations of physical systems can or cannot be brought about with unbounded accuracy by devices capable of operating in a cycle ('constructors'). Hence, in particular, such specifications cannot refer to time. Thus, although laws expressed in constructor-theory-compatible form automatically avoid the anomalous properties of time in traditional formulations of fundamental theories, this very feature raises the problem of how they can nevertheless give meaning to duration and dynamics, and thereby be consistent with traditionally formulated laws. Here we show how.

## 1. Introduction: why time is problematic

In the *traditional conception* of laws of physics, time appears as a real-valued parameter on which physical quantities depend. This dependence is typically expressed in the form of differential equations in which time is an independent variable. Yet this variable is not itself a physical quantity in the usual sense. For example, in quantum mechanics, it is not a quantum observable but a c-number, commuting with all quantum observables, including all possible Hamiltonians. Hence it has no dynamics and cannot be measured. Nevertheless it is supposed to 'change', passing through a range of values. In general relativity, time is part of spacetime – which does have dynamics, yet the problem arises in a different way: an observer cannot measure proper time except by reference to clocks which have their own dynamics. A consistent theory of the dynamics of time would therefore have to be 'timeless'. That is to say, time would not be an independent background parameter but rather a phenomenon that emerges, under suitable circumstances and perhaps only approximately, from relationships between physical systems (see e.g. Barbour 2012 , Isham 1992).

Under constructor theory, fundamental laws of physics do not refer to time. They are expressed as timeless statements about which transformations could or could not be brought about with unbounded accuracy by devices capable of operating in a cycle with unbounded reliability. We call such devices *approximate constructors*. *Constructors* are their idealisations to perfect accuracy and reliability (see Section 3). Catalysts and heat engines are familiar approximate constructors. Transformations that can be so brought about are called *possible* and those that cannot are called *impossible.* We call laws consisting only of such statements *constructor-theory-compatible*.

Hence, in regard to time, constructor theory seems to have the mirror-image of the traditional problem we noted above: since time does not appear in constructor-theory-compatible laws, how can change, dynamics and duration be expressed in constructor-theoretic terms? In this paper we explain how.

Our approach shares some features with other 'timeless' approaches to time (Wheeler 1968, Dewitt 1967, Page & Wootters 1983, Rovelli 2004). As in those approaches, it regards time as having no existence independently of physical objects, and as existing only as relationships between their attributes such as clock readings. But unlike those approaches, ours is independent of specific theories of matter and spacetime.

## 2. Inaccuracy, unreliability and limits

Constructor theory requires perfect determinism in the following specific sense: that whatever transformation a given device would effect on a given object in a given state, it would also effect it on any other object of the same constitution in the same state. So, in particular, probability, which requires multiple outcomes to be possible, has no exact physical meaning under constructor theory. A constructor-theory-compatible statement of, say, a quantum uncertainty principle (e.g. Bialynicki-Birula & Mycielski 1975), must therefore be expressed in terms of reliably bringing about transformations on variables expressed via density operators, not on probability distribution functions. Just as one of us (Marletto 2015) has given sufficient conditions for a constructor-theory-compatible theory to support the applicability of the probability calculus, here we do the same for time.

The reader may notice that here and in what follows, we are not using exclusively constructor-theoretic language to explain what constructor theory is. Any attempt to do so would be irrelevant here (as well as circular). For this paper is about how to express, in constructor-theory-compatible terms, *laws of physics*, and in particular laws that are currently expressed in terms of a time 'variable'.

As explained in (Violaris & Marletto, 2022), given a system that could perform a given (pairwise) task, *accuracy* is defined as the distance between the desired output attribute of the task, and the actual output of the system being used to perform it, when run once;

*reliability* is the degree to which such a system retains its ability to deliver the task to that accuracy, when re-used repeatedly. Which measure of distance to use in these cases is up to the specific constructor-theory-compatible theory to specify. For instance, in (Violaris & Marletto, 2022), this is quantified with a measure called 'relative deterioration' expressed in terms of the trace distance defined on quantum theory's Hilbert spaces. Constructor theory requires as a necessary condition for a theory to be constructor-theory-compatible, merely that such a measure exists.

We are not aiming for an exact characterisation of time, but only an emergent one. So we shall make two simplifying assumptions from the outset. The first is that no device can operate perfectly accurately or reliably. So, exact constructor-theory-compatible laws must be expressed as statements of whether there is or is not a bound, short of perfection, on how accurately and reliably each putative transformation can be brought about, and on how well the device that brings it about can operate in a cycle. If there is such a bound, we call the transformation *impossible*. Otherwise we call it 'possible' – meaning only that there is an infinite sequence of possible devices that could bring about the transformation with increasing accuracy and reliability, with perfection as their least upper bound.

The sequence of transformations that those devices would bring about would converge to the specified transformation (convergence being relative to some measure of deviation from perfection). But the devices themselves – approximate constructors – need not converge to anything. To avoid cumbersome terminology, we use the term *constructor*[1] for a fictitious ideal device that could bring about a given transformation perfectly and repeatedly.

## 3. Possible and impossible tasks

An *attribute* of a physical system is a set of some of the system's possible states. Laws of physics under constructor theory are expressed in terms of *tasks*. In this paper it will suffice to consider a *task* to be an ordered pair of attributes $x$ and $y$ of a physical system $\mathcal{P}$, which we write as

[1] The term 'constructor' in this sense was introduced by von Neumann (von Neumann, 1966), but in an abstract (not explicitly physical) model of the logic of biological reproduction, which later became the theory of cellular automata.

$$x \rightarrow y \text{ on } \mathcal{P}. \qquad (1)$$

$\mathcal{P}$ with attribute $x$ is called the *input* of the task (1), and $\mathcal{P}$ with attribute $y$ is the *output.* No other meaning is given to the terms 'input' and 'output', except that the performance of a task must produce not only a specified output but also some *measurable indication* that the transformation is complete – just as a Turing machine must be capable of indicating that its computation has 'halted'. (Measurement and computation both have purely constructor-theoretic meanings as defined in Deutsch & Marletto (2015).) The indication may be an attribute of either the constructor or the output. If it were not for that requirement, then, for instance, a computer (being a type of constructor) would qualify as being capable of deciding an undecidable question given as input, simply by displaying *yes,* then *no,* repeatedly.

We shall refer to any physical system on which tasks are given meanings by laws of physics, as a *substrate*. In situations where the laws of physics allow two substrates $\mathcal{P}_1$ and $\mathcal{P}_2$ with attributes $x$ and $y$ to be jointly an input of a task, we can regard them as a single, composite substrate $\mathcal{P}_1 \oplus \mathcal{P}_2$ with attribute $(x, y)$.

If there is no finite bound on the accuracy or reliability with which a task $X$ can be performed by approximate constructors, then $X$ is possible, and a constructor for $X$ is too – a *possible substrate* being one whose existence is not forbidden by the laws of physics. If $X = (x \rightarrow y \text{ on } \mathcal{P})$, we write both these statements of possibility as $X^{\checkmark}$, i.e.

$$(x \rightarrow y \text{ on } \mathcal{P})^{\checkmark}. \qquad (2)$$

Otherwise the task is *impossible*, which we write $X^{\times}$ or

$$(x \rightarrow y \text{ on } \mathcal{P})^{\times}. \qquad (3)$$

Note that 'possible' does not mean 'happens sometimes' nor 'happens' at all. Statements such as (2) are counterfactual in character; but determinism is required, so there is no third option between possible and impossible. It also follows that there is no constructor-theory-compatible law specifying initial conditions for the universe. They must be emergent consequences of the laws of physics, just as the final state is under the prevailing conception of laws of physics.

In addition, a task does not specify what must happen *during* its performance, nor how long a constructor would take to perform it – the latter being the origin of the problem of time in constructor theory, which we solve here.

If a set $\boldsymbol{\mathcal{P}} = \{\mathcal{P}_1, \mathcal{P}_2 \dots\}$ of substrates are each regarded as having possible attributes $x$ and $y$, then we can specify a task $x \to y$ on $\boldsymbol{\mathcal{P}}$. This is defined as possible if a single constructor is capable of accepting any $\mathcal{P} \in \boldsymbol{\mathcal{P}}$ with attribute $x$ as input, and of transforming it so as to have attribute $y$. Note that

$$(x \to y \text{ on } \boldsymbol{\mathcal{P}})^{\checkmark} \tag{4}$$

is a more onerous condition than

$$(\forall \mathcal{P} \in \boldsymbol{\mathcal{P}})((x \to y \text{ on } \mathcal{P})^{\checkmark}), \tag{5}$$

because (4) requires the *same constructor* to be capable of effecting $x \to y$ on any $\mathcal{P} \in \boldsymbol{\mathcal{P}}$, without being informed which is it being given, while (5) would hold if a different constructor could do so for each $\mathcal{P} \in \boldsymbol{\mathcal{P}}$. A straightforward example from quantum theory is that if for each unit vector $\boldsymbol{n}$ in three dimensions, $\mathcal{P}_{\boldsymbol{n}}$ is a spin-based qubit with computation basis elements 0 and 1 in the directions of the unit vectors $\pm\boldsymbol{n}$, and $F$ is the task of reversing the direction of a qubit along that axis, and $\boldsymbol{\mathcal{P}}$ is the set of all such qubits, then

$$(\forall \boldsymbol{n})((F \text{ on } \mathcal{P}_{\boldsymbol{n}})^{\checkmark}), \text{ yet } (F \text{ on } \boldsymbol{\mathcal{P}})^{\text{✗}}, \tag{6}$$

since the latter task is forbidden by the laws of quantum theory: it would violate unitarity.

In addition to the hope that laws of physics similar or equivalent to known ones can be formulated in constructor-theory-compatible terms, as well as ones with new content, constructor theory has overarching laws of physics in its own right. These are laws about laws: physical *principles.* For a summary of conjectured principles of constructor theory, see (Marletto &  2013, Deutsch & Marletto 2014, Marletto 2020). Here we shall use the *composition principle*: it requires the composition of possible tasks to be possible tasks (Deutsch 2013). We shall also rely implicitly on the *principle of locality*, in the following constructor-theory-compatible form (Deutsch 2013, Deutsch & Marletto 2015), modelled on that of Einstein (1949):

> There exists a description where attributes of substrates are such that any attribute of a composite substrate $\mathcal{A} \oplus \mathcal{B}$ is an ordered pair of attributes $(\boldsymbol{a}, \boldsymbol{b})$ of substrate $\mathcal{A}$ and substrate $\mathcal{B}$,with the property that performing any task on $\mathcal{A}$ only, cannot change the attribute $\boldsymbol{b}$ of $\mathcal{B}$.

(We have here introduced the symbol $\oplus$ to denote the composition of two distinct substrates.) If this principle of locality holds, the same will hold for states, since a set of one state is by definition an attribute. Note that here by "state" we mean the complete specification of the factual state of affairs of a given substrate, not necessarily restricted to what is empirically accessible by measuring observables of that system only. This is why this principle is satisfied by both relativistic and non-relativistic quantum theory (notwithstanding Bell non-locality – see (Deutsch & Hayden, 2000)).

## 4. Isolated substrates and static attributes

The foundation of the theory of time in constructor theory is the notion of an *isolated substrate*. In the traditional conception of physics, an isolated system is one whose dynamical interactions do not depend on variables of other systems. In constructor theory, we state this property without referring to time or dynamics, in the following way. A substrate has the attribute of being *isolated* if the possibility of any task on it alone depends only on its own attributes. It follows, by definition, that the composite substrate of a number of isolated substrates is also isolated. Constructor theory does not require that all substrates can exist in (or arbitrarily close to) isolated form, but it does require that the combined system of a constructor and its substrates can. It also requires that all laws of physics can expressed in terms of tasks. This is implicit in our notation and descriptions of physical laws in terms of possibilities and impossibilities of tasks. Note also that the possibility of isolated substrates is implied by the locality principle, in the form stated in section 3.

The next notion is that of a static attribute. In the traditional conception, a *static variable* is one that does not change as long as the substrate is isolated. It commutes with the system's intrinsic Hamiltonian and hence is unaffected by time. How can the staticity of $x$ be expressed under constructor theory? Not like this:

$$(x \to y \text{ on } \mathcal{P})^{\text{✗}} \quad (y \neq x), \qquad (7)$$

because that would not require $\mathcal{P}$ to be isolated from the constructor. Moreover, a constructor that performs (7) by interacting with $\mathcal{P}$ *will* typically exist, provided that no conservation law is broken. Indeed, (7) expresses that $x$ is an attribute where a conserved quantity is sharp, not the stronger condition that $x$ be static. For example, $x$ and $y$ could be distinct degenerate eigenstates of the free Hamiltonian of $\mathcal{P}$ and the constructor could transform one to the other adiabatically, but the prohibition (7) holds if $x$ and $y$ are not degenerate.

A *static* attribute is one that it is impossible to change if the substrate is isolated. That is in view of the fact the only constructor that could 'act' on it would be a timer – as we shall now discuss.

## 5. The null task and constructors for it

Consider the *null task* { }, which is represented by the empty set. This task has no constraints on what substrate it is to be performed, nor does it have constraints on its allowed input or required outputs. It is therefore the most lenient task. Remarkably, the composition principle implies that the null task must be possible, $\{\ \}^{\checkmark}$. For, upon denoting via the symbol '•' the serial composition of two tasks, the null task can be written as the serial composition of two possible tasks, whose inputs and outputs are distinct from each other:

$$\{\boldsymbol{a} \to \boldsymbol{b}\} \bullet \{\boldsymbol{c} \to \boldsymbol{d}\} = \{\quad\}, \quad (\boldsymbol{b} \cap \boldsymbol{c} = \emptyset)\ .$$

Hence, by the composition principle, it must be a possible task. A constructor for that task, we shall call a *null constructor*.

As we shall now argue, a *null constructor* has all the properties that would, in the traditional conception, qualify it as a timer, and its possibility is necessary for the existence of time in the traditional sense. It is not a *sufficient* condition because constructor theory needs to be compatible with exotic variants of time such as closed timelike lines and spacetime foam. Constructor-theory-compatible theories are permitted to make sense of those or rule them out.

Given that the null task has no substrate, a null constructor $C$ for it doesn't act on one and must therefore be *an isolated substrate* itself. This, remarkably, implies that a null constructor $C$ must have at least three attributes with specific properties. First, since { } specifies no output, the measurable indication of completion must be an attribute of $C$ itself. Let us call it $\mathbf{1}$. To work as a halt flag, $\mathbf{1}$ must in turn be distinguishable[2] from another attribute $\boldsymbol{R}$ of $C$, meaning 'running'. There must also be a third attribute, $\mathbf{0}$, meaning 'starting' – this attribute must be preparable if $\{\ \}^{\checkmark}$. That is to say, $C$, isolated with attribute $\mathbf{0}$, transforms itself to having attribute $\boldsymbol{R}$, and, having attribute $\boldsymbol{R}$, transforms itself to having attribute $\mathbf{1}$. Evidently $\mathbf{0}$ and $\boldsymbol{R}$ cannot be static attributes but $\mathbf{1}$ must be, or it would fail to indicate 'completed' correctly. Interestingly, having these three attributes gives the null constructor the characteristic features of a particular kind of clock – a *timer*, which we shall now define in constructor-theory-compatible terms.

## 6. Timers

In the traditional conception, a *timer with duration* $t$ is a device that can be used to determine whether a physical process lasts more or less as long as the fixed interval $t$ characteristic of the timer. An idealised example in the traditional conception is an infinite straight line labelled by integers a distance $d$ apart, along which a classical point particle moves with constant velocity $d/t$. The attribute $\mathbf{0}$ corresponds to the particle being at position 0. It is there only for an instant: $\mathbf{0}$ is a non-static attribute. The non-static attribute $\boldsymbol{R}$ corresponds to its being strictly between positions 0 and 1, and the static attribute $\mathbf{1}$ to its being at position 1 or beyond.[3] If timers defined under constructor theory are to resemble arbitrarily closely the timers that one can describe in the traditional conception of physics, there must be a 1-parameter family of sets $\boldsymbol{C}_t$, each containing all possible timers of duration $t$. Here we are thinking of the very simplest type of timer, whose output is not numerical, but rather is a single bit that the user can measure, after the timer has been prepared with attribute $\mathbf{0}$, to determine whether the period $t$ since the timer was prepared has ended (attribute $\mathbf{1}$) or not (attribute $\boldsymbol{R}$). Thus a *timer* in this sense consists of: a null constructor with a particular choice of attributes $\mathbf{0}$, $\mathbf{1}$, $\boldsymbol{R}$ and another attribute, the (raised) halt flag. The duration $t$ of the timer is implicit once these attributes are specified. In general, the halt flag is a (sub-) substrate of which the raised and non-raised halt flag are attributes, which must be distinguishable from each other. The halt flag cannot always be chosen to coincide with the attribute $\mathbf{1}$, as the latter may not belong to a set of distinguishable states, while the halt flag must belong to such a set.

---

[2] 'Distinguishable' can be defined in a dynamics-independent way, using only constructor-theoretic concepts, in terms of copiable variables, (Deutsch&Marletto, 2015). See the appendix in Marletto, 2022, for a summary of these concepts.

[3] Because of Poincaré recurrence, no attribute of a finite physical system can be strictly static, nor, under reversible dynamics, can a non-static attribute spontaneously evolve to a static one. But these properties can be approximated arbitrarily closely in practice. For example, consider a counter with a finite number $N$ of binary digits, so it can count from 0 to $2^N - 1$ before it cycles. If the counter starts at 0, it will eventually show 0 again. But for moderately large $N$, this will never happen in practice.

Now we use the properties of isolated systems and the definitions given so far to define a few properties for timers, which defines an equivalence class on the set of all timers, such that each class is labelled by a particular duration $t$ for all the timers in the class. Consider any composite substrate $C_t \oplus C_{t'}$ consisting of timers from the sets $\boldsymbol{C}_t$ and $\boldsymbol{C}_{t'}$ respectively, with $t' > t > 0$. This composite substrate constitutes another timer in the set $\boldsymbol{C}_t$ if it is defined as halting with the attribute $(\mathbf{1}, \boldsymbol{R})$ and never with any other attribute: in particular not $(\mathbf{1}, \mathbf{1})$ and by definition not $(\boldsymbol{R}, \mathbf{1})$. This means that it will halt when $C_t$ does, so it belongs to the same set of timers. Hence we shall denote the composite system $C_t \oplus C_{t'}$, when defined as another timer belonging to $C_t$ , with the symbol $[C_t \oplus C_{t'}]_t$.

The properties of isolated substrates imply that:

$$\left((\mathbf{0}, \mathbf{0}) \rightarrow (\mathbf{1}, \mathbf{1}) \text{ on } [C_t \oplus C_{t'}]_t\right)^{\boldsymbol{\mathsf{X}}} \tag{8}$$

whenever $t' > t > 0$. Property (8) follows from the fact that if $t' > t$, the sets $\boldsymbol{C}_t$ and $\boldsymbol{C}_{t'}$ have no elements in common. Hence if a timer from $\boldsymbol{C}_t$ and one from $\boldsymbol{C}_{t'}$ both start having the non-static attribute ($\mathbf{0}$,$\mathbf{0}$), they cannot transform to the attribute ($\mathbf{1}$,$\mathbf{1}$), given that each of the two substrates is isolated and, so, the possibility of tasks performable on each of them depends on their attributes only. This rules out any constructor tampering with either $\boldsymbol{C}_t$ or $\boldsymbol{C}_{t'}$ to make them faster or slower.

On the other hand, when $t' = t$, the two timers in $C_t \oplus C_{t'}$ belong both to the same set $\boldsymbol{C}_t$; hence by definition they both transform to the attribute $\mathbf{1}$ and signal they have halted, while each of them is isolated, in time $t$. So we have:

$$\left((\mathbf{0},\mathbf{0}) \rightarrow (\mathbf{1},\mathbf{1}) \text{ on } [\mathcal{C}_t \oplus \mathcal{C}_{t'}]_t\right)^{\checkmark} \qquad (9)$$

if and only if $t=t'$.

It follows that the constructor for the possible task (9) is itself a timer of duration $t$. But that information is only implicit in (9). A constructor-theoretic specification of a task must not mention the constructor. It also follows that whenever (9) is true for two timers $\boldsymbol{\mathcal{C}}_t$ and $\boldsymbol{\mathcal{C}}_{t'}$, then these two timers belong to the same equivalence class, which we can interpret as the class of all timers with the same duration $t$.

A particular substrate may obey those conditions for more than one choice of the attributes $\mathbf{0}$, $\mathbf{1}$, $\boldsymbol{R}$ and the halt flag. Such a substrate, including the specification of those choices, would therefore appear more than once in $\boldsymbol{\mathcal{C}}_t$ for a given $t$.

We say that a set of constructor-theory-compatible laws of physics *allows for timers* to the extent that there are sets $\boldsymbol{\mathcal{C}}_t$ of isolated substrates with attributes $\mathbf{0}$, $\mathbf{1}$, $\boldsymbol{R}$ and the halt flag, for which the conditions (8) and (9) hold.

Given the properties of isolated substrates, any pair of isolated identical instances $\mathcal{C} \oplus \mathcal{C}$ of a timer $\mathcal{C}$ with joint attribute $(\boldsymbol{x},\boldsymbol{x})$ cannot transform itself to any attribute whose two constituent attributes are not identical – i.e. not of the form $(\boldsymbol{y},\boldsymbol{y})$. This implements another of the traditional-conception properties of time: the synchrony of mutually isolated clocks. Noting how this synchrony is implemented in the traditional conception – particularly in general relativity – we infer that the state of motion of a substrate must potentially appear as an attribute, and the spacetime region it exists in must appear as an additional substrate, in constructor-theory-compatible laws of physics for that substrate. The same must therefore go for constructor-theory-compatible laws governing spacetime itself, if and when general relativity is recast in constructor-theory-compatible form.

## 7. Dynamics

In order to recover dynamics as in the traditional conception, one needs to identify the law specifying how given variables of an isolated substrate $\mathcal{P}$ change relative to the standard labels of a variable of a class of timers. Specifically, to express differential equations for physical variables in constructor-theory-compatible terms, we define (say) a real variable as a set $V = \{v(\lambda)\}$ of disjoint non-static attributes $v(\lambda)$ of an isolated substrate $\mathcal{P}$, indexed by a parameter $\lambda$. (In quantum theory, these attributes could be q-numbers; but note that in general they need not be observables.) Then one considers the composite substrate $\mathcal{P} \oplus \mathcal{C}_{\Delta\lambda}$, where $\mathcal{C}_{\Delta\lambda}$ is a timer with duration $\Delta\lambda$ and the attributes $x = v(\lambda)$ and $x' = v(\lambda + \Delta\lambda)$ such that:

$$\left((\boldsymbol{x}, \mathbf{0}) \rightarrow (\boldsymbol{x}', \mathbf{1}) \text{ on } \mathcal{P} \oplus \mathcal{C}_{\Delta\lambda}\right)^{\checkmark} \tag{10}$$

The differential equations expressing the dynamical evolution of the variable V of $\mathcal{P}$ is obtained considering the incremental ratio:

$$\frac{\mathrm{d}v}{\mathrm{d}\lambda} = \frac{v(\lambda + \Delta\lambda) - v(\lambda)}{\Delta\lambda} \tag{11}$$

and taking the limit $\Delta\lambda \rightarrow 0$.

In particular, this procedure allows us to express the dynamical evolution of a clock $\mathcal{P}$ with some discrete or continuous pointer variable.

## 8. Relation to 'timeless' theories of time in the traditional conception

The peculiarities of time that we mentioned in Section 1 have inspired some physicists to seek 'timeless' formulations of the laws of physics – ones in which no time parameter appears explicitly and only the states of clocks do. There are such formulations for quantum theory (Page & Wootters 1983, Smith 2021, Kuypers 2022, Rijavec 2023), for approaches to quantum gravity (Wheeler & DeWitt, DeWitt 1967, Ashtekar 2007, Kiefer 2007, Rovelli 2004), and even for classical dynamics (Barbour 2012). The constructor theory of time that we have presented here is also 'timeless', in that its principles do not explicitly refer to time; and time and dynamics are explained in relation to timers and clocks. However, unlike in the existing timeless approaches, we do not rely on the formalisms and properties of existing dynamics; hence this theory offers an explanatory foundation for such approaches themselves.

Timeless approaches to time in the traditional conception assume a number of properties of natural laws – such as the possibility of good clocks; the principle of locality; and the possibility of superinformation media (generalisations of quantum systems, as defined in (Deutsch and Marletto, 2014). In this paper we have expressed, as conjectured laws of physics, regularities in nature that are necessary for timeless approaches to time to be possible. Further conditions may be imposed by particular theories: in the traditional conception, they are expressed as relationships between variables expressing the state of the universe. In constructor theory, they must be expressed as requirements about possible and impossible tasks; and may involve the requirements that superinformation media are possible. Finally, we note that theories that conform to the principle of constructor theory but do not support timers are not ruled out by the conditions we have set out.

## 9. Conclusions

We have proposed a theory of clocks and timers in constructor theory, providing conditions for timer-supporting theories that are constructor-theory compliant. Then we have shown how in such theories one could recover dynamics as emerging from the timeless principles of constructor theory and the theory of clocks, specifically timers. This work provides the foundations for timeless approaches to dynamics, rooting them in general, timeless principles of physics.

**Acknowledgements**

The authors thank Sam Kuypers and Vlatko Vedral for several helpful comments on earlier versions of this manuscript. CM thanks the Gordon and Betty Moore Foundation, the Eutopia Foundation and B. Vass for supporting her research.